\documentclass[final,3p,times]{elsarticle}

\usepackage{amssymb}
\usepackage{amsmath}
\usepackage{myphi}

\journal{Journal of Subatomic Particles and Cosmology}

\newcommand{\sNNn}[1]{\sNN $=$ \SI{#1}{\GeV}}

\graphicspath{{picts/}}

\begin{document}

\begin{frontmatter}

\title{News from \NASixtyOne}

\author[aaa]{Antoni Marcinek}
\ead{antoni.marcinek@ifj.edu.pl}
\author{for the \NASixtyOne Collaboration}
\affiliation[aaa]{organization={Henryk Niewodniczański Institute of Nuclear Physics, Polish Academy of Sciences},
             addressline={Radzikowskiego 152},
             city={Kraków},
             postcode={31-342},
             country={Poland}}

\begin{abstract}
\NASixtyOne is a multi-purpose, fixed-target hadron spectrometer at the CERN
SPS. Its research program includes studies of strong interactions, as well as
reference measurements for neutrino and cosmic-ray physics. A significant
advantage of \NASixtyOne over collider experiments is its extended coverage of
phase space available for particle production.
This contribution summarizes recent selected results from our strong interaction
program. These include rapidity spectra of \Kp and \pim mesons in \XeLa
reactions compared to other collision systems, energy dependence of the scaled
\pim total yield (the \enquote{kink} plot) in \XeLa compared to other collision
systems and models, \phiM rapidity spectra in \ArSc collisions compared to
models, $\Lambda$ transverse polarization in \pp reactions, energy dependence of
intensive quantities of net-electric charge distribution in \ArSc collisions
compared to \pp results, as well as an update on the evidence of large violation
of isospin symmetry in kaon production --- the excess of charged over neutral
kaons.
\end{abstract}

\begin{keyword}
strangeness production \sep hadron production \sep critical point \sep
\NASixtyOne \sep isospin-symmetry breaking \sep relativistic heavy-ion
collisions
\end{keyword}

\end{frontmatter}

\section{Introduction}
\begin{figure}[tb]
  \centering
  \includegraphics[width=\textwidth]{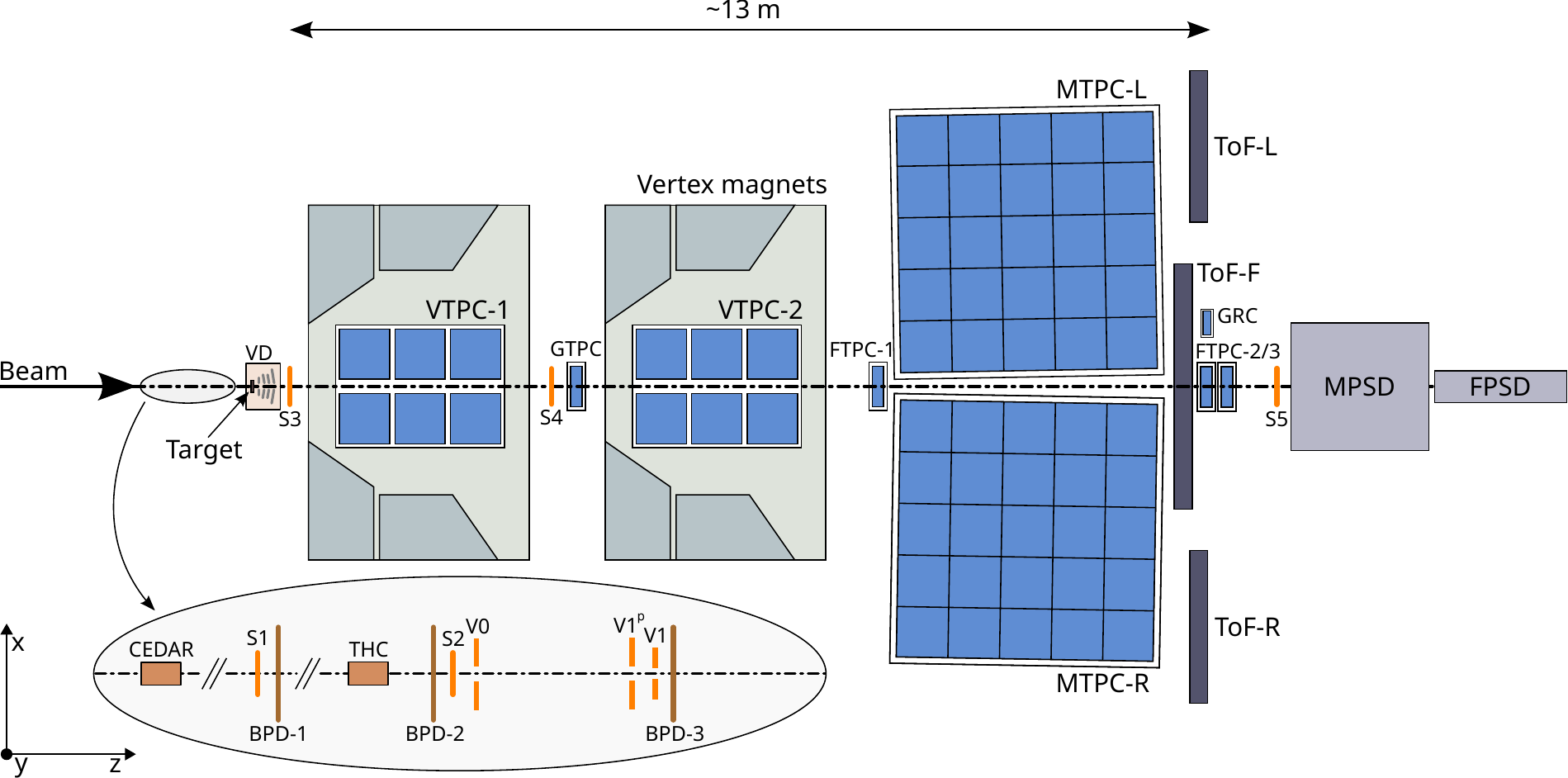}
  \caption{Schematic layout of the \NASixtyOne detector system (horizontal cut
    in the beam plane, not to scale) showing the state of the detector as of
    autumn 2025.
  }
  \label{fig:det}
\end{figure}
\begin{figure}[p]
  \centering
  \includegraphics[width=\textwidth,clip, trim=0 0 25mm 0]{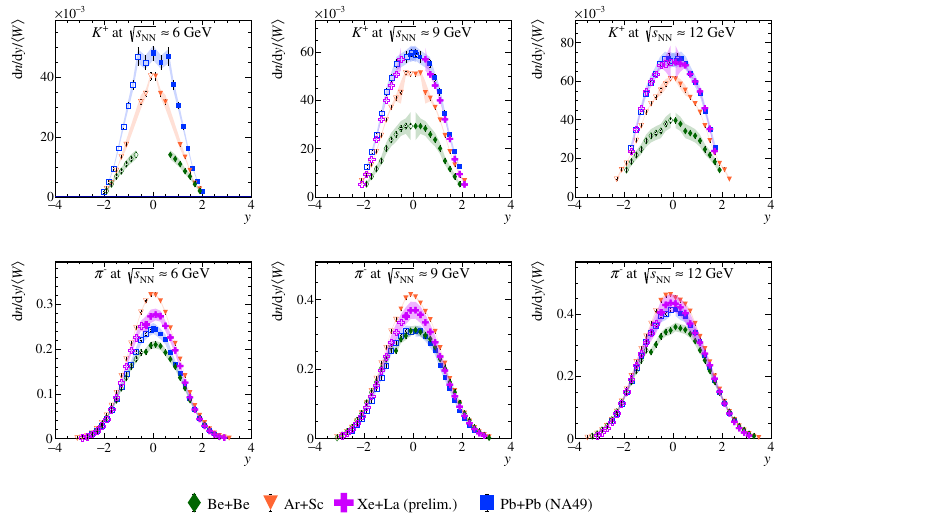}
  \caption[]{Comparison of rapidity distributions normalized by the mean number of
    wounded nucleons for positively-charged kaons (top) and negatively-charged
    pions (bottom) produced in central \BeBe~\capcite{NA61SHINE:2020czq,
    NA61SHINE:2020ggt}, \ArSc~\capcite{NA61SHINE:2023epu, NA61SHINE:2021nye},
    \XeLa,
    and \PbPb~\capcite{NA49:2002pzu, NA49:2007stj} collisions at three collision
    energies indicated in the plots. Statistical uncertainties are shown with
    error bars, systematic uncertainties are shown as shaded bands.
    For all spectra, open points are reflected measured data points.
  }
  \label{fig:dndyW}
  \vspace{2em}
  \includegraphics[width=\textwidth,clip, trim=9.6mm 31.5mm 9.6mm 13mm]{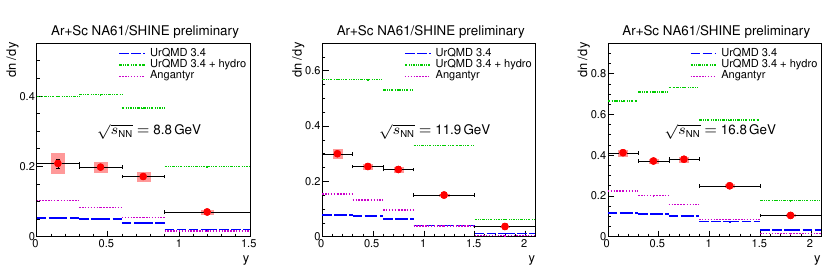}
  \caption{Rapidity distributions of \phiM mesons produced in the 10\% most
    central \ArSc collisions measured by the \NASixtyOne
    experiment (red circles), compared to model predictions.
    Vertical bars represent statistical uncertainties, red rectangles systematic
    uncertainties, and horizontal bars depict rapidity bin sizes.
    Model calculations were performed by S.~Veli (Technical University of Munich) and
    T.~Janiec (The University of Manchester).
  }
  \label{fig:dndy}
\end{figure}
The \NASixtyOne detector is a multi-purpose, fixed-target hadron spectrometer at
the CERN SPS. Its schematic layout is shown in \cref{fig:det}. Its main
components are large-volume Time Projection Chambers, two of which are immersed
in a magnetic field perpendicular to the beam, allowing for momentum
measurement. Such configuration gives \NASixtyOne a significant advantage over
collider experiments in the studies of particle production: very large
acceptance. Longitudinally, the entire forward hemisphere of the collision is
covered for charged hadrons, while for neutral hadrons decaying into a pair of
charged particles, additionally a large part of the backward hemisphere is
accessible. In the transverse direction, the experiment has no low-\pt cut-off.
This is accompanied by excellent particle identification capabilities via the
energy loss in TPCs (\dEdx), as well as time-of-flight measurements. Collision
centrality is measured with forward energy deposits
in two hadronic calorimeters (Projectile Spectator Detectors).
\par
The above allows for a rich, two-fold physics program: studies of strong
interactions, as well as reference precision measurements for neutrino and
cosmic-ray physics. This contribution focuses on the former, which includes the
study of the onset of deconfinement, search for the critical point of strongly
interacting matter and recently measurements of the open charm and
isospin-symmetry violation in multiparticle production.

\section{Study of the onset of deconfinement}
The onset of deconfinement~\cite{Gazdzicki:2010iv} corresponds to an energy threshold separating hadron
production dominated by formation and decays of hadron resonances to that
dominated by creation and hadronization of Quark Gluon Plasma (QGP). It is
studied in \NASixtyOne in the first two-dimensional scan in the history of
heavy-ion collisions, in the collision energy and the colliding system size.
\Cref{fig:dndyW} shows some of the results of that scan: a comparison of
rapidity distributions normalized by the colliding system size expressed as the
mean number of wounded nucleons~$\expval{W}$ for two selected particle species. Our new
preliminary results in the central \XeLa collisions are compared to that in
central \BeBe~\cite{NA61SHINE:2020czq, NA61SHINE:2020ggt},
\ArSc~\cite{NA61SHINE:2023epu, NA61SHINE:2021nye} and \PbPb~\cite{NA49:2002pzu,
NA49:2007stj} reactions. One can see that for \Kp there is a monotonic
dependence of the scaled yield on the system size, presumably associated with
strangeness enhancement, further discussed in \recite{LewickiSQM2026}. On the
other hand, \pim production exhibits a non-monotonic dependence of the scaled
yield on the system size, most striking at midrapidity, a phenomenon which is
currently not understood.
\par
\NASixtyOne studies strangeness enhancement also in the hidden-strangeness
sector, \ie for \phiM mesons. Their rapidity distributions for the 10\% most
central \ArSc collisions are compared to model predictions at three collision
energies in \cref{fig:dndy}. These are the first-ever results on \phiM
production in intermediate-size systems at CERN SPS. It is clear that none of
the considered models matches the experimental data points. It was also shown
previously~\cite{Rozplochowski:2024pld} that the $\phi/\pi$ ratio for \ArSc
collisions is comparable to that in \PbPb reactions, and significantly higher
than for \pp interactions.
\begin{figure}[t]
  \centering
  \includegraphics[width=\textwidth,clip, trim=20mm 40mm 20mm 0]{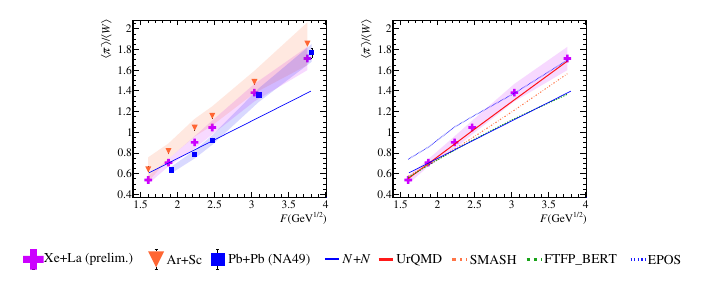}
  \vspace{-3em}
  \caption[]{Fermi variable (energy) dependence of the $\expval{\pi^-}/\expval{W}$
    ratio (the \enquote{kink} plot) for pions produced in central \XeLa,
    \ArSc~\capcite{NA61SHINE:2021nye} and  \PbPb~\capcite{NA49:2002pzu, NA49:2007stj}
    collisions, compared to several models indicated in the legend, as well as
    parametrization of nucleon-nucleon collisions~\capcite{Panova:2025caw}.
  }
  \label{fig:kink}
  \vspace{2em}
  \includegraphics[width=0.6\textwidth]{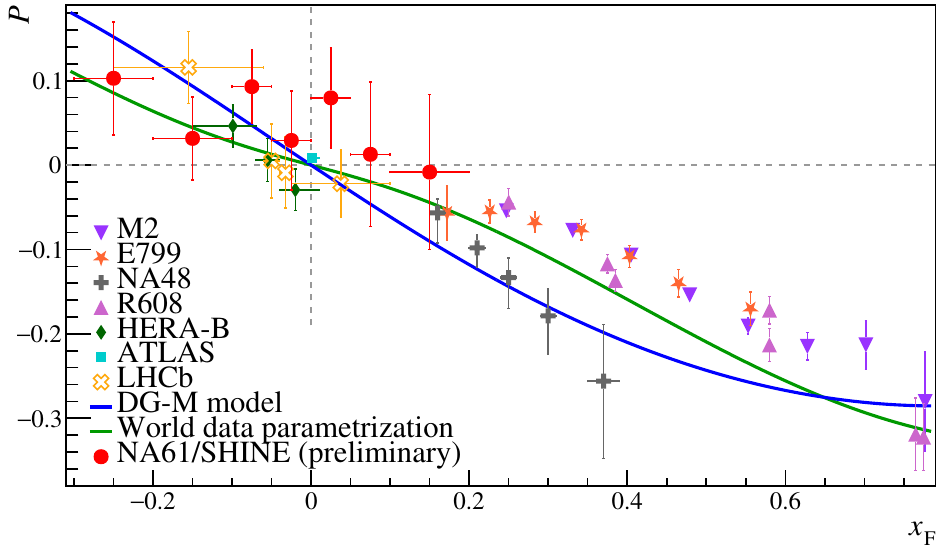}
  \caption[]{$\Lambda$ transverse polarization measured by \NASixtyOne in \pp
    collisions at \sNNn{17.3} for $\pt \in (0.8, 1.2)$ \si{\GeVc}, compared to
    the world proton-proton and proton-nucleus data~\capcite{Lundberg:1989hw,
    Ramberg:1994tk, Fanti:1998px, R608:1986ltk, HERA-B:2006rds, ATLAS:2014ona,
    LHCb:2025rxf} at $\pt \approx \SI{1}{\GeVc}$ and the  DeGrand-Miettinen model
    prediction~\capcite{DeGrand:1980gc}. Total uncertainties are plotted.
  }
  \label{fig:lambdaP}
\end{figure}
\begin{figure}[t]
  \centering
  \includegraphics[width=0.33\textwidth]{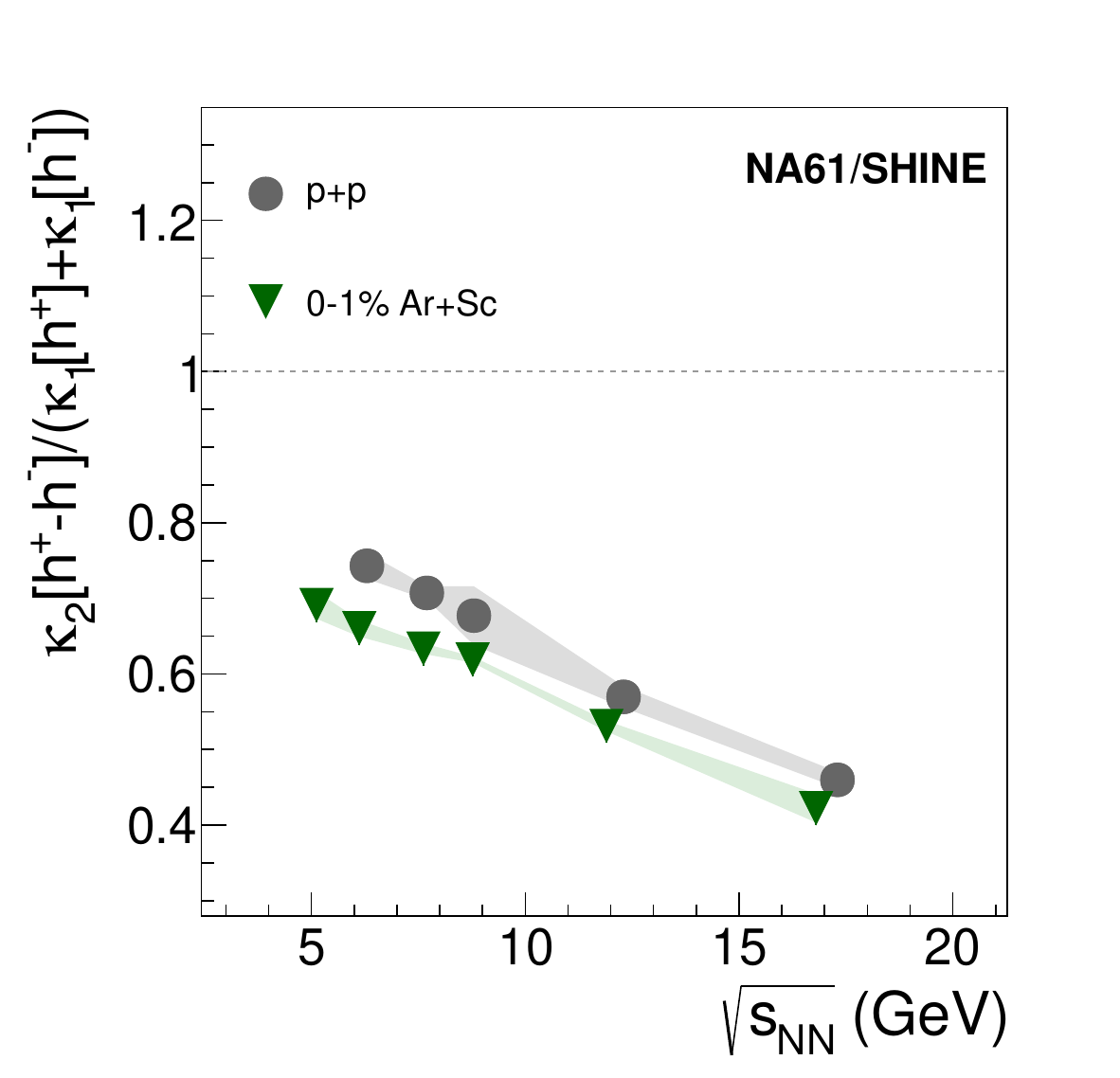}
  \includegraphics[width=0.33\textwidth]{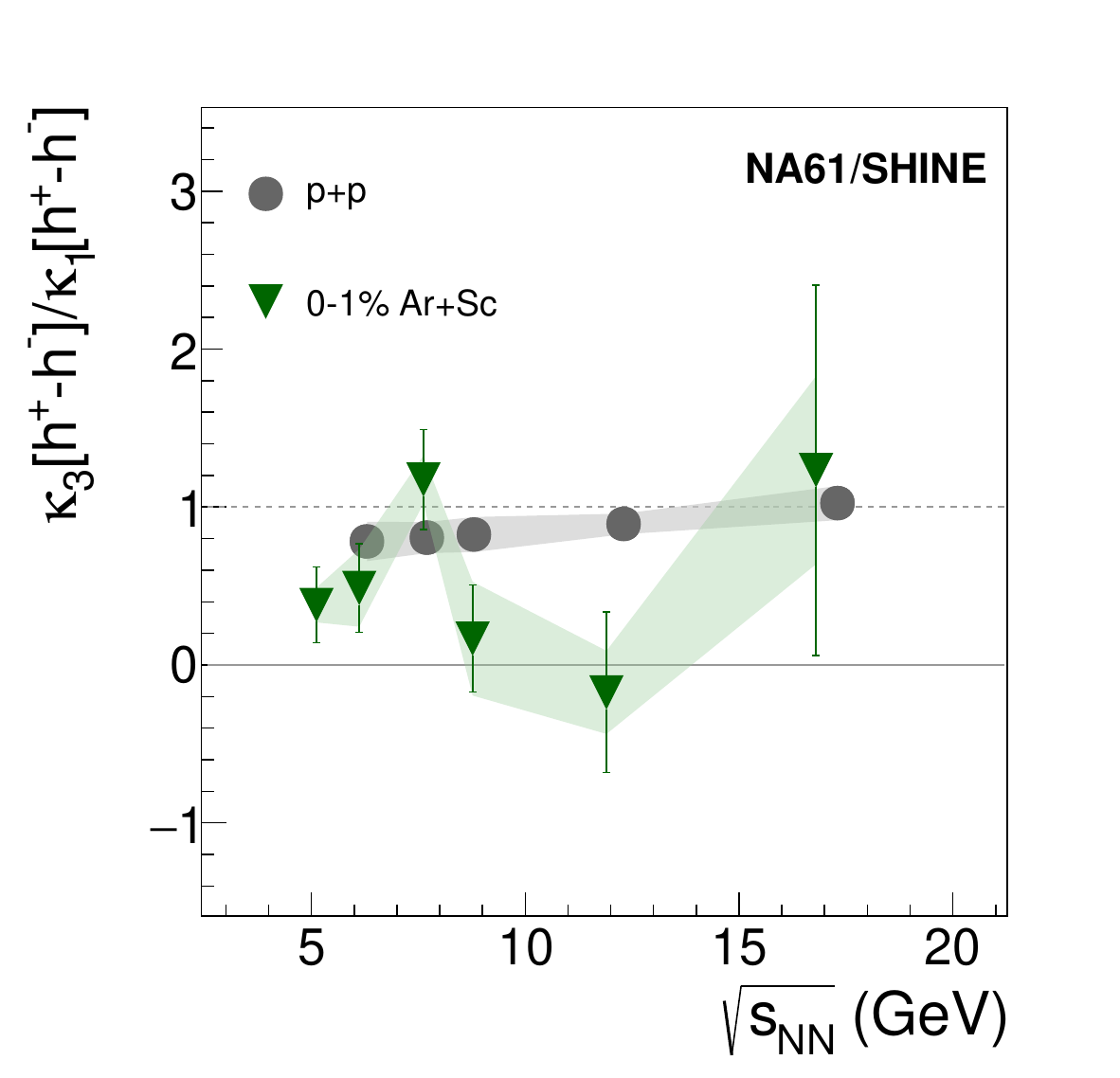}
  \includegraphics[width=0.33\textwidth]{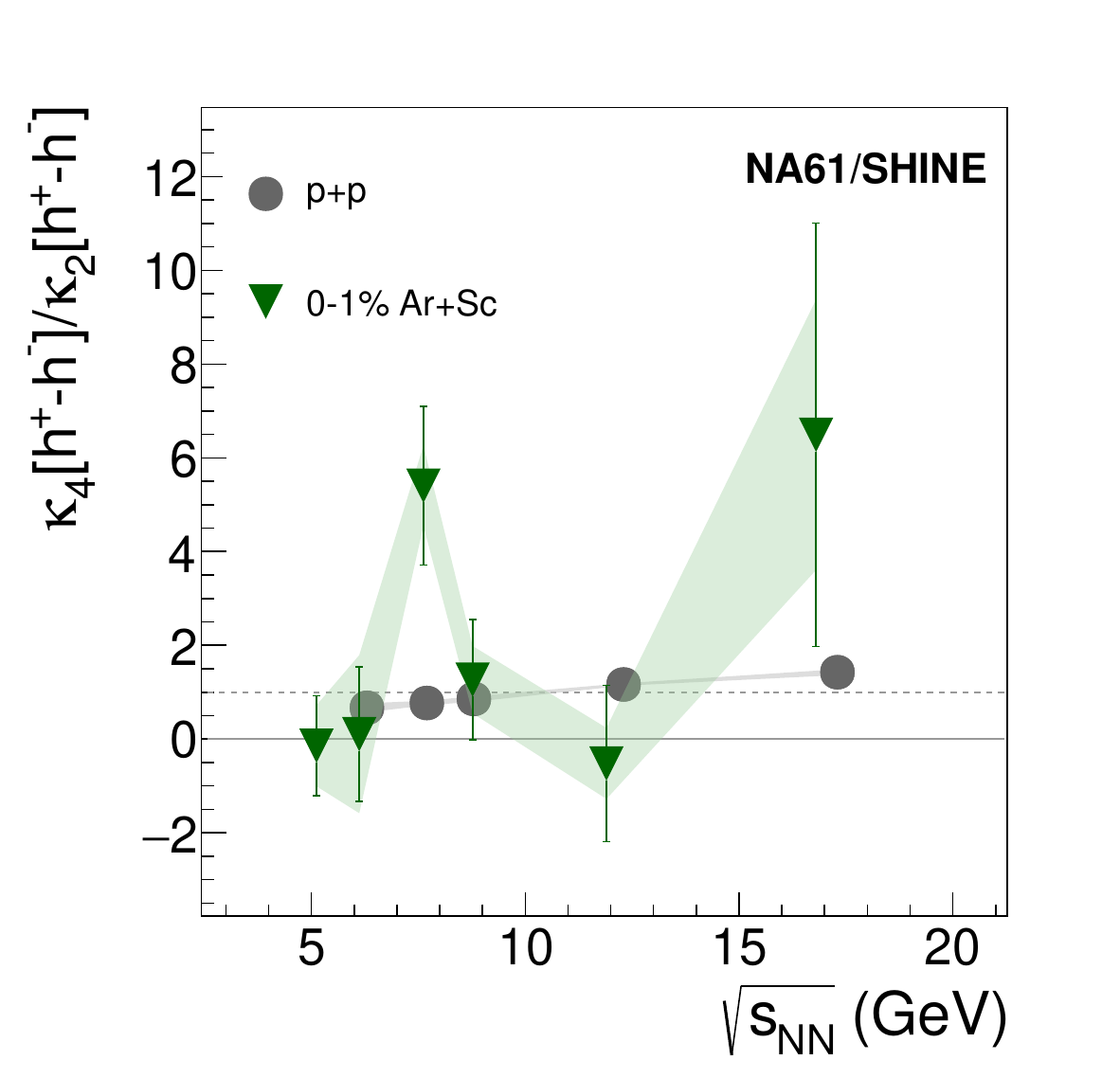}
  \caption[]{Energy dependence of intensive quantities of net-electric charge
    distribution in \ArSc collisions~\capcite{NA61SHINE:2025whi} compared to \pp
    results~\capcite{NA61SHINE:2023unn}. Statistical uncertainties are shown with
    error bars, systematic uncertainties are shown as shaded bands.
  }
  \label{fig:cumulants}
\end{figure}
\par
One of observables proposed~\cite{Gazdzicki:1998vd} as a signal of the onset of
deconfinement is the enhancement of the entropy production visible as a \emph{kink} in
the plot of entropy over system size as a function of the Fermi variable $F =
\frac{(\sNN - 2m_\text{N})^{3/4}}{\sNN^{1/4}}$, where $m_\text{N}$ is the
nucleon mass. Because at CERN SPS energies \pim are
by far the most abundant particles, they can serve as a proxy for the measure of
entropy. \Cref{fig:kink} presents our preliminary results on the Fermi variable
dependence of the $\expval{\pim}/\expval{W}$ ratio,
the \pim total yield scaled by the mean number of wounded nucleons,
in the central \XeLa collisions, compared to \ArSc~\cite{NA61SHINE:2021nye} and
\PbPb~\cite{NA49:2002pzu, NA49:2007stj} data, several models, and
parametrization of nucleon-nucleon collisions~\cite{Panova:2025caw}. In \XeLa
collisions a hint of a kink is visible, similar to that in \PbPb data. The
dependence in nucleus-nucleus collisions is clearly significantly steeper than
that for the parametrization of the nucleon-nucleon data. Comparing to models,
it is visible that only \UrQMD successfully describes \XeLa results, while for
other considered models the dependence is less steep.
\par
Another of possible observables associated with the deconfinement transition is
system-size dependence of the $\Lambda$ transverse
polarization~\cite{Panagiotou:1989sv}. \Cref{fig:lambdaP} presents preliminary
\NASixtyOne results on the $\Lambda$ polarization in \pp collisions at
\sNNn{17.3}. Given substantial uncertainties, they are consistent with zero
polarization at midrapidity ($x_F=0$) and agree with the world proton-proton and
proton-nucleus data~\cite{Lundberg:1989hw, Ramberg:1994tk, Fanti:1998px,
  R608:1986ltk, HERA-B:2006rds, ATLAS:2014ona, LHCb:2025rxf} and the
  DeGrand-Miettinen model prediction~\cite{DeGrand:1980gc}.

\section{Search for the critical point of strongly interacting matter}
With the data from the two-dimensional scan, \NASixtyOne carries out various
correlation and fluctuation studies in the search for the critical point of
strongly interacting matter. \Cref{fig:cumulants} presents energy dependence of
cumulant ratios of net-electric charge distribution in the central
\ArSc collisions~\cite{NA61SHINE:2025whi}. It is compared to \pp
results~\cite{NA61SHINE:2023unn} which serve as the non-critical baseline.
Cumulants are defined according to:
\begin{align*}
  \kappa_1 & = \langle N \rangle \,, \\
  \kappa_2 & = \langle(\delta N)^2 \rangle = \sigma^2 \,, \\
  \kappa_3 & = \langle(\delta N)^3 \rangle = S\sigma^3 \,, \\
  \kappa_4 & = \langle(\delta N)^4 \rangle - 3\langle(\delta N)^2 \rangle^2 =
  K\sigma^4 \,,
  \label{eq:cum}
\end{align*}
where $N$ is multiplicity, $\delta N = N - \langle N \rangle$, $\sigma$ is the
standard deviation, $S$ is the skewness, and $K$ is the kurtosis. Cumulant ratios are
used to form intensive quantities which are volume-independent, although they
are still sensitive to volume fluctuations. To reduce the latter effect, only
the 1\% most central \ArSc collisions are considered. For the cumulant ratios
certain reference values are indicated in the plots: in the case of no
fluctuations, the ratios should be equal to zero; in the case of independent
particles, governed by the Skellam distribution, the ratios should be equal to
one. In general the measurements follow neither of these reference values. A
hint of non-monotonic behaviour is visible for higher order cumulants $\kappa_3/\kappa_1$ and
$\kappa_4/\kappa_2$ in \ArSc collisions, but uncertainties are too large to
draw firm conclusions. This non-monotonicity might be related to the critical
point~\cite{Stephanov:1998dy}, but also to the non-critical phenomenon of the
onset of deconfinement~\cite{Gazdzicki:2026ubm}.

\section{Unexpected excess of charged over neutral kaons}
\begin{figure}[tb]
  \centering
  \includegraphics[width=0.7\textwidth]{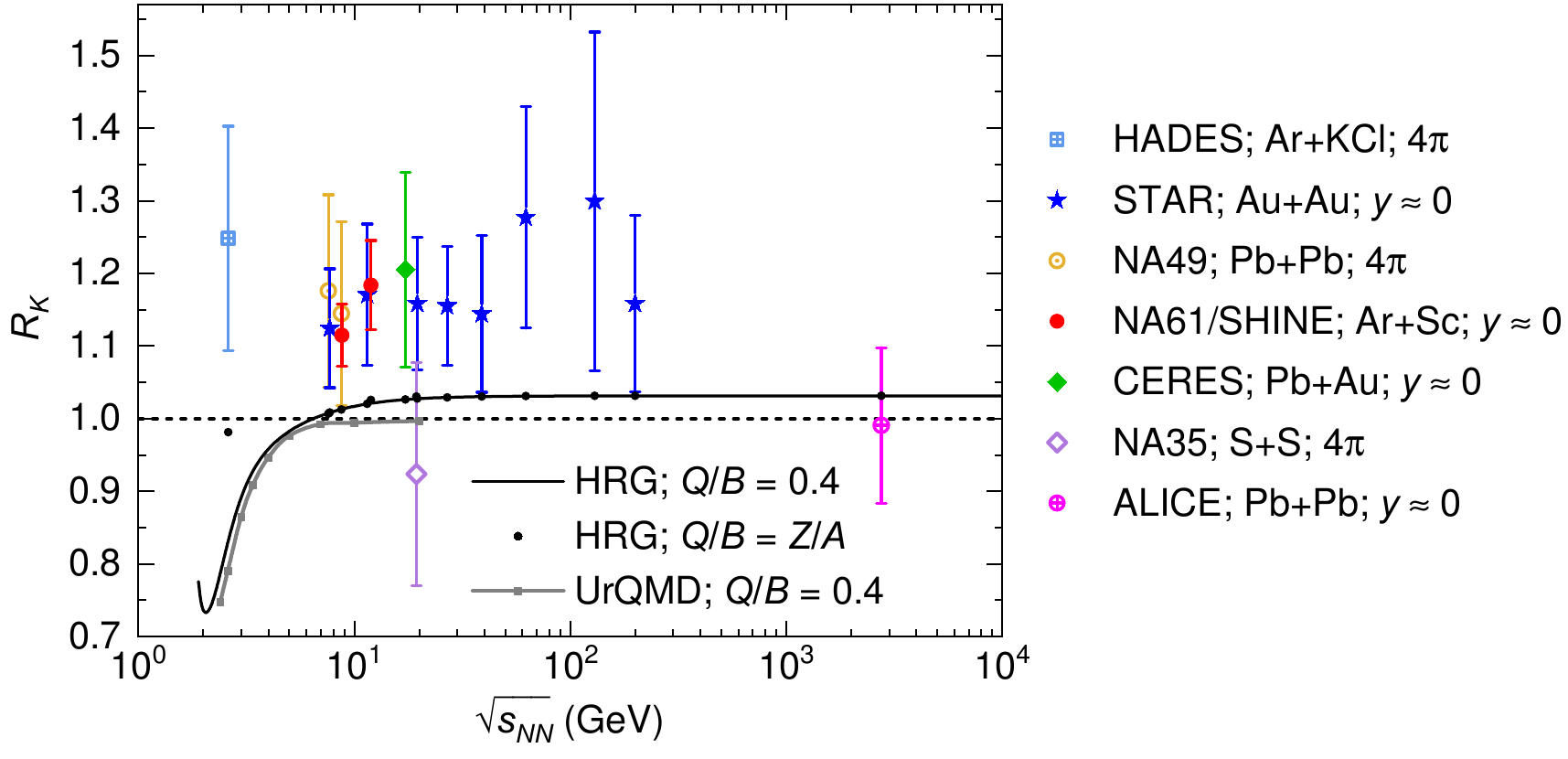}
  \caption[]{Energy dependence of the $R_K$ ratio (see text) measured by
    \NASixtyOne and other experiments~\capcite{NA61SHINE:2023azp} and new
    preliminary \NASixtyOne point for \ArSc at \sNNn{8.8}, in 4$\pi$ or
    midrapidity depending on the experiment, compared to model predictions.
  }
  \label{fig:isospin}
\end{figure}
An unexpected  excess of charged over neutral kaons was first observed by
\NASixtyOne within the cosmic-ray physics part of the program, in the \pimC
collisions~\cite{NA61SHINE:2022tiz}. It was not reproduced by the models nor
quark counting~\cite{NA61SHINE:2022tiz}. Due to peculiarity of the \pimC system,
at that time it was difficult to interpret the result. Then came measurements in
the nearly charge-symmetric \ArSc collisions and it became possible to give the
observation a meaning.
For exact isospin symmetry, \ie equality of $u$ and $d$ quark masses and collisions of
charge-symmetric ($Z=N$) nuclei, one expects for the yields of hadrons:
$\expval*{\Kp}(u\bar{s})=\expval*{K^0}(d\bar{s})$
and
$\expval*{\Km}(\bar{u}s)=\expval*{\overline{K}^0}(\bar{d}s)$.
Neglecting small CP violation we also have
$\expval*{K_S^0} = \frac{1}{2}\expval*{K^0} +
    \frac{1}{2}\expval*{\overline{K}^0} = \expval*{K_L^0}$.
Therefore in this case, the expected charged to neutral kaon ratio:
\begin{equation}
  R_K
    = \frac{\expval*{\Kp} + \expval*{\Km}}{\expval*{K^0} + \expval*{\overline{K}^0}}
    = \frac{\expval*{\Kp} + \expval*{\Km}}{2 \expval*{K_S^0}} = 1 \,.
  \label{eq:isospin}
\end{equation}
The energy dependence of this ratio measured by \NASixtyOne in \ArSc collisions
at \sNNn{8.8} and \sNNn{11.9}, together with values resulting from a compilation
of measurements of \Kp, \Km, and \KOS from other experiments~\cite{NA61SHINE:2023azp}, is shown in
\cref{fig:isospin}. It is compared to the expectation from models reflecting
known effects of the isospin-symmetry violation~\cite{NA61SHINE:2023azp}. The
measured ratios are clearly higher than that predicted by the models. Taking
into account the new preliminary \NASixtyOne result for \ArSc collisions at
\sNNn{8.8}, the significance of the isospin-symmetry violation beyond the
known effects is 5.3$\sigma$.

\section{Summary}
\NASixtyOne measured a rich collection of spectra in the two-dimensional collision
energy and system-size scan, studying the onset of deconfinement. We have
observed non-monotonic dependence of the $\expval{\pi^-}/\expval{W}$ ratio on the system size,
similarity of the energy dependence of the $\expval{\pi^-}/\expval{W}$ ratio in \XeLa and \PbPb
collisions, and inability of the considered models to describe \phiM production
in \ArSc collisions. We measured no clear signal of the critical point across
various fluctuation and correlation studies. We observe an unexpected excess of
charged over neutral kaon production in \pimC and \ArSc collisions, which is an
indication of violation of isospin symmetry beyond the known effects in \ArSc
reactions.\\[2ex]
This work was supported by the National Science Centre, Poland (grant number
2023\slash 51\slash D\slash ST2\slash 02950).

\bibliographystyle{elsarticle-num}
\bibliography{bibliography}

@misc{LewickiSQM2026,
    author = "Balkova, Y. and Lewicki, M.",
    eprint = "2608.04758",
    archivePrefix = "arXiv",
    primaryClass = "nucl-ex",
    note = "these proceedings"
}

@article{Gazdzicki:1998vd,
    author = "Gaździcki, Marek and Gorenstein, Mark I.",
    OPTtitle = "{On the early stage of nucleus-nucleus collisions}",
    OPTeprint = "hep-ph/9803462",
    archivePrefix = "arXiv",
    OPTreportNumber = "IKF-HENPG-2-98",
    journal = "Acta Phys. Polon. B",
    volume = "30",
    pages = "2705",
    year = "1999"
}

@article{Gazdzicki:2010iv,
    author = "Gazdzicki, M. and Gorenstein, M. and Seyboth, P.",
    OPTtitle = "{Onset of deconfinement in nucleus-nucleus collisions: Review for pedestrians and experts}",
    OPTeprint = "1006.1765",
    archivePrefix = "arXiv",
    primaryClass = "hep-ph",
    OPTdoi = "10.5506/APhysPolB.42.307",
    journal = "Acta Phys. Polon. B",
    volume = "42",
    pages = "307",
    year = "2011"
}

@article{NA61SHINE:2020czq,
    author = "Acharya, A. and others",
    collaboration = "NA61/SHINE",
    OPTtitle = "{Measurements of $\pi^\pm$, $K^\pm$, $p$ and $\bar{p}$ spectra in $^7$Be+$^9$Be collisions at beam momenta from 19$A$ to 150$A$ GeV/$c$ with the NA61/SHINE spectrometer at the CERN SPS}",
    OPTeprint = "2010.01864",
    archivePrefix = "arXiv",
    primaryClass = "hep-ex",
    reportNumber = "CERN-EP-2020-187, CERN-EP-2020-187",
    OPTdoi = "10.1140/epjc/s10052-020-08733-x",
    journal = "Eur. Phys. J. C",
    volume = "81",
    OPTnumber = "1",
    pages = "73",
    year = "2021",
    note = "[Erratum: Eur.Phys.J.C 83, 90 (2023)]"
}

@article{NA61SHINE:2023epu,
    author = "Adhikary, H. and others",
    collaboration = "NA61/SHINE",
    OPTtitle = "{Measurements of $\pi ^\pm $, $K^\pm $, p and $\bar{p}$ spectra in $^{40}\hbox {Ar+}^{45}\hbox {Sc}$ collisions at 13A to 150A~$\text{ Ge }\hspace{-1.00006pt}\text{ V }\!/\!c$}",
    OPTeprint = "2308.16683",
    archivePrefix = "arXiv",
    primaryClass = "nucl-ex",
    reportNumber = "CERN-EP-2023-179, FERMILAB-PUB-23-563-AD",
    OPTdoi = "10.1140/epjc/s10052-024-12602-2",
    journal = "Eur. Phys. J. C",
    volume = "84",
    OPTnumber = "4",
    pages = "416",
    year = "2024"
}

@article{NA61SHINE:2020ggt,
    author = "Acharya, A. and others",
    collaboration = "NA61/SHINE",
    OPTtitle = "{Measurements of $\pi^-$ production in $^7$Be+$^9$Be collisions at beam momenta from 19$A$ to 150$A$GeV/$c$ in the NA61/SHINE experiment at the CERN SPS}",
    OPTeprint = "2008.06277",
    archivePrefix = "arXiv",
    primaryClass = "nucl-ex",
    reportNumber = "CERN-EP-2020-150",
    OPTdoi = "10.1140/epjc/s10052-020-08514-6",
    journal = "Eur. Phys. J. C",
    volume = "80",
    OPTnumber = "10",
    pages = "961",
    year = "2020",
    note = "[Erratum: Eur.Phys.J.C 81, 144 (2021)]"
}

@article{NA61SHINE:2021nye,
    author = "Acharya, A. and others",
    collaboration = "NA61/SHINE",
    OPTtitle = "{Spectra and mean multiplicities of $\pi ^{-}$ in central${}^{40}$Ar+${}^{45}$Sc collisions at 13A, 19A, 30A, 40A, 75A and 150$A\,\text{ Ge }\text{ V }\!/\!\textit{c}$ beam momenta measured by the NA61/SHINE spectrometer at the CERN SPS}",
    OPTeprint = "2101.08494",
    archivePrefix = "arXiv",
    primaryClass = "hep-ex",
    reportNumber = "CERN-EP-2021-010",
    OPTdoi = "10.1140/epjc/s10052-021-09135-3",
    journal = "Eur. Phys. J. C",
    volume = "81",
    OPTnumber = "5",
    pages = "397",
    year = "2021"
}

@article{NA49:2002pzu,
    author = "Afanasiev, S. V. and others",
    collaboration = "NA49",
    OPTtitle = "{Energy dependence of pion and kaon production in central Pb + Pb collisions}",
    OPTeprint = "nucl-ex/0205002",
    archivePrefix = "arXiv",
    OPTdoi = "10.1103/PhysRevC.66.054902",
    journal = "Phys. Rev. C",
    volume = "66",
    pages = "054902",
    year = "2002"
}

@article{NA49:2007stj,
    author = "Alt, C. and others",
    collaboration = "NA49",
    OPTtitle = "{Pion and kaon production in central Pb + Pb collisions at 20-A and 30-A-GeV: Evidence for the onset of deconfinement}",
    OPTeprint = "0710.0118",
    archivePrefix = "arXiv",
    primaryClass = "nucl-ex",
    OPTdoi = "10.1103/PhysRevC.77.024903",
    journal = "Phys. Rev. C",
    volume = "77",
    pages = "024903",
    year = "2008"
}

@phdthesis{Panova:2025caw,
    author = "Panova, Oleksandra",
    OPTtitle = "{CHARGED HADRON PRODUCTION IN CENTRAL Xe+La COLLISIONS AT THE CERN SPS}",
    OPTdoi = "10.17181/f9r5b-na547",
    school = "Jan Kochanowski University",
    year = "2025"
}

@article{Lundberg:1989hw,
    author = "Lundberg, B. and others",
    OPTtitle = "{Polarization in Inclusive $\Lambda$ and $\bar{\Lambda}$ Production at Large $p_T$}",
    reportNumber = "FERMILAB-PUB-89-269-E",
    OPTdoi = "10.1103/PhysRevD.40.3557",
    journal = "Phys. Rev. D",
    volume = "40",
    pages = "3557",
    year = "1989"
}

@article{Ramberg:1994tk,
    author = "Ramberg, E. J. and others",
    OPTtitle = "{Polarization of $\Lambda$ and $\bar{\Lambda}$ Produced by 800 GeV Protons}",
    reportNumber = "FERMILAB-PUB-94-015",
    OPTdoi = "10.1016/0370-2693(94)91397-8",
    journal = "Phys. Lett. B",
    volume = "338",
    pages = "403",
    year = "1994"
}

@article{Fanti:1998px,
    author = "Fanti, V. and others",
    OPTtitle = "{A Measurement of the transverse polarization of Lambda hyperons produced in inelastic p N reactions at 450-GeV proton energy}",
    reportNumber = "PRINT-99-001",
    OPTdoi = "10.1007/s100520050337",
    journal = "Eur. Phys. J. C",
    volume = "6",
    pages = "265",
    year = "1999"
}

@article{R608:1986ltk,
    author = "Smith, A. M. and others",
    collaboration = "R608",
    OPTtitle = "{$\Lambda^0$ Polarization in Proton Proton Interactions From $sqrt{s} = 31$-{GeV} to 62-{GeV}}",
    reportNumber = "CERN-EP/86-182",
    OPTdoi = "10.1016/0370-2693(87)91556-5",
    journal = "Phys. Lett. B",
    volume = "185",
    pages = "209",
    year = "1987"
}

@article{HERA-B:2006rds,
    author = "Abt, I. and others",
    collaboration = "HERA-B",
    OPTtitle = "{Polarization of Lambda and anti-Lambda in 920-GeV fixed-target proton-nucleus collisions}",
    OPTeprint = "hep-ex/0603047",
    archivePrefix = "arXiv",
    reportNumber = "DESY-06-027",
    OPTdoi = "10.1016/j.physletb.2006.05.040",
    journal = "Phys. Lett. B",
    volume = "638",
    pages = "415",
    year = "2006"
}

@article{ATLAS:2014ona,
    author = "Aad, Georges and others",
    collaboration = "ATLAS",
    OPTtitle = "{Measurement of the transverse polarization of $\Lambda$ and $\bar{\Lambda}$ hyperons produced in proton-proton collisions at $\sqrt{s}=7$ TeV using the ATLAS detector}",
    OPTeprint = "1412.1692",
    archivePrefix = "arXiv",
    primaryClass = "hep-ex",
    reportNumber = "CERN-PH-EP-2014-258",
    OPTdoi = "10.1103/PhysRevD.91.032004",
    journal = "Phys. Rev. D",
    volume = "91",
    OPTnumber = "3",
    pages = "032004",
    year = "2015"
}

@article{LHCb:2025rxf,
    author = "Aaij, Roel and others",
    collaboration = "LHCb",
    OPTtitle = "{Measurement of transverse {\ensuremath{\Lambda}} and {\ensuremath{\Lambda}}{\textasciimacron} hyperon polarization in pPb collisions at sNN=5.02{\,}{\,}TeV}",
    OPTeprint = "2508.02009",
    archivePrefix = "arXiv",
    primaryClass = "nucl-ex",
    reportNumber = "LHCb-PAPER-2025-004, CERN-EP-2025-153",
    OPTdoi = "10.1103/rc6r-zt9q",
    journal = "Phys. Rev. D",
    volume = "112",
    OPTnumber = "11",
    pages = "112022",
    year = "2025"
}

@article{NA61SHINE:2025whi,
    author = "Adhikary, H. and others",
    collaboration = "NA61/SHINE",
    OPTtitle = "{Multiplicity and net-electric charge fluctuations in central Ar+Sc interactions at 13A, 19A, 30A, 40A, 75A, and 150$A\,\hbox {GeV}\!/\!c$ beam momenta measured by NA61/SHINE at the CERN SPS}",
    OPTeprint = "2503.22484",
    archivePrefix = "arXiv",
    primaryClass = "nucl-ex",
    reportNumber = "CERN-EP-2025-056, FERMILAB-PUB-25-0245-AD",
    OPTdoi = "10.1140/epjc/s10052-025-14621-z",
    journal = "Eur. Phys. J. C",
    volume = "85",
    OPTnumber = "8",
    pages = "918",
    year = "2025"
}

@article{NA61SHINE:2023unn,
    author = "Adhikary, H. and others",
    collaboration = "NA61/SHINE",
    OPTtitle = "{Measurements of higher-order cumulants of multiplicity and net-electric charge distributions in inelastic proton-proton interactions by NA61/SHINE}",
    OPTeprint = "2312.13706",
    archivePrefix = "arXiv",
    primaryClass = "hep-ex",
    OPTdoi = "10.1140/epjc/s10052-024-13076-y",
    journal = "Eur. Phys. J. C",
    volume = "84",
    pages = "921",
    year = "2024",
    note = "[Erratum: Eur.Phys.J.C 85, 341 (2025)]"
}

@article{NA61SHINE:2023azp,
    author = "Adhikary, H. and others",
    collaboration = "NA61/SHINE",
    OPTtitle = "{Evidence of isospin-symmetry violation in high-energy collisions of atomic nuclei}",
    OPTeprint = "2312.06572",
    archivePrefix = "arXiv",
    primaryClass = "nucl-ex",
    reportNumber = "CERN-EP-2023-283",
    OPTdoi = "10.1038/s41467-025-57234-6",
    journal = "Nature Commun.",
    volume = "16",
    OPTnumber = "1",
    pages = "2849",
    year = "2025"
}

@article{DeGrand:1980gc,
    author = "DeGrand, Thomas A. and Miettinen, Hannu I.",
    OPTtitle = "{Quark Dynamics of Polarization in Inclusive Hadron Production}",
    reportNumber = "UCSB-TH-24-1980, HU-TFT-80-47",
    OPTdoi = "10.1103/PhysRevD.23.1227",
    journal = "Phys. Rev. D",
    volume = "23",
    pages = "1227",
    year = "1981"
}

@article{Rozplochowski:2024pld,
    author = "Rozp{\l}ochowski, {\L}ukasz",
    collaboration = "NA61/SHINE",
    OPTtitle = "{Energy dependence of {\ensuremath{\phi}}(1020) meson production in nucleus-nucleus collisions at the CERN SPS}",
    OPTeprint = "2410.02379",
    archivePrefix = "arXiv",
    primaryClass = "nucl-ex",
    OPTdoi = "10.1051/epjconf/202531605003",
    journal = "EPJ Web Conf.",
    volume = "316",
    pages = "05003",
    year = "2025"
}

@article{Panagiotou:1989sv,
    author = "Panagiotou, Apostolos D.",
    OPTtitle = "{$\Lambda^0$ Polarization in Hadron - Nucleon, Hadron - Nucleus and Nucleus-nucleus Interactions}",
    reportNumber = "CERN-EP-89-131",
    OPTdoi = "10.1142/S0217751X90000568",
    journal = "Int. J. Mod. Phys. A",
    volume = "5",
    pages = "1197",
    year = "1990"
}

@article{Stephanov:1998dy,
    author = "Stephanov, Misha A. and Rajagopal, K. and Shuryak, Edward V.",
    OPTtitle = "{Signatures of the tricritical point in QCD}",
    OPTeprint = "hep-ph/9806219",
    archivePrefix = "arXiv",
    reportNumber = "ITP-SB-98-39, MIT-CTP-2748, SUNY-NTG-98-17",
    OPTdoi = "10.1103/PhysRevLett.81.4816",
    journal = "Phys. Rev. Lett.",
    volume = "81",
    pages = "4816",
    year = "1998"
}

@article{Gazdzicki:2026ubm,
    author = "Gazdzicki, Marek and Gorenstein, Mark and Rustamov, Anar",
    OPTtitle = "{Baryon fluctuation signatures of the onset of deconfinement}",
    eprint = "2603.15819",
    archivePrefix = "arXiv",
    primaryClass = "nucl-th",
    year = "2026"
}

@article{NA61SHINE:2022tiz,
    author = "Adhikary, H. and others",
    collaboration = "NA61/SHINE",
    OPTtitle = "{Measurement of hadron production in \ensuremath{\pi}--C interactions at 158 and 350\,\,GeV/c with NA61/SHINE at the CERN SPS}",
    OPTeprint = "2209.10561",
    archivePrefix = "arXiv",
    primaryClass = "nucl-ex",
    OPTdoi = "10.1103/PhysRevD.107.062004",
    journal = "Phys. Rev. D",
    volume = "107",
    OPTnumber = "6",
    pages = "062004",
    year = "2023"
}

\end{document}